**DTIP 2007** of MEMS & MOEMS  Stresa, Italy, 25-27 April 2007# NOISE AND THERMAL STABILITY OF VIBRATING MICRO-GYROMETERS PREAMPLIFIERS

*R. Levy[1], A. Dupret[1], H. Mathias[1], J-P. Gilles[1], Fabien Parrain[1], Bruno Eisenbeis[2], S. Megherbi[1]*

[1]ONERA, 29 avenue de la division Leclerc
92320 Châtillon

[2]IEF, bat. 220, université Paris sud
91405 Orsay**ABSTRACT**

*The preamplifier is a critical component of gyrometer's electronics. Indeed the resolution of the sensor is limited by its signal to noise ratio, and the gyrometer's thermal stability is limited by its gain drift. In this paper, five different kinds of preamplifiers are presented and compared. Finally, the design of an integrated preamplifier is shown in order to increase the gain stability while reducing its noise and size.*## 1. INTRODUCTION

During the last decades, vibrating MEMS have been widely developed worldwide and have met many applications such as inertial sensors: accelerometers [1,2] and gyrometers [3,4], bio-sensors [5], or force microscopy[6]. The mechanical parameter to sense (amplitude or frequency variation) is converted into charges by capacitive sensing or piezoelectricity. These small charges are then sensed with preamplifiers.

For the gyrometer, the goal of the first detection stage is to sense the amplitude of the Coriolis induced charges on the detection electrodes. This voltage is then converted into a dc voltage proportional to the angular velocity with analog or digital electronics. As the detected charges are very small compared to the parasitic charges induced by parasitic mechanical and capacitive couplings [7], the preamplifier is a critical component. Indeed its signal to noise ratio limits the resolution of the sensor, and the thermal stability of the gain limits the thermal stability of the sensor's output.

A few kinds of preamplifiers have already been studied like the differential charge amplifier [8,9] and the switched capacitor charge amplifier [10,11]. But these architectures have not been compared, and their thermal stability has not been taken into account.

In this paper, we will compare five different preamplifiers in terms of noise and thermal stability. The benefits and drawbacks of each kind of circuit are discussed. The best suited circuit for the gyro is integrated in 0.35μm BiCMOS technology, and specific design and layout is performed to ensure low noise and thermal stability

## 2. SPECIFICATIONS FOR THE GYRO'S PREAMPLIFIER

### 2.1. Description of the gyro's associated electronics

The physical phenomenon used for vibrating gyros is the Coriolis force induced by rotation. The shape of the gyro, either a ring [12], a tuning fork [13] or a disk [14], allows two orthogonal modes of vibration; the drive mode along the x axis which is excited at resonance, and the detection mode along the y axis induced by the Coriolis force due to a rotation along the z axis. The amplitude of the detection mode is proportional to the angular rate velocity.

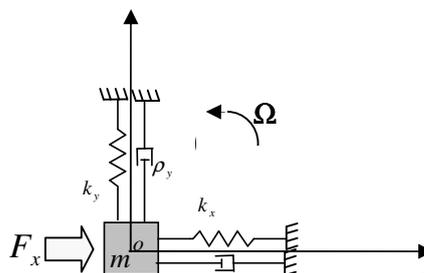

Figure 1: mass and stiffness model of a vibrating micro-gyro. Ω is the angular velocity and Fx the drive force.

The drive mode is excited by the voltage Vx at its resonance frequency $\omega_x$, and the charges on the electrodes of the detection mode $q_y+$ and $q_y-$ include the Coriolis charges, the charges induced by capacitive coupling, and those induced by mechanical coupling:

$$V_X = X \cos(\omega_x t + \varphi) \quad \text{(eq. 5)}$$

$$q_y+ = \underbrace{\varepsilon(\Omega).\sin(\omega_x t)}_{\text{Coriolis signal}} + \underbrace{C\sin(\omega_x t)}_{\text{Capacitive coupling}} + \underbrace{M\cos(\omega_x t)}_{\text{Mechanical coupling}} \quad \text{(eq. 6)}$$

$$q_y- = \underbrace{-\varepsilon(\Omega).\sin(\omega_x t)}_{\text{Coriolis signal}} + \underbrace{C\sin(\omega_x t)}_{\text{Capacitive coupling}} - \underbrace{M\cos(\omega_x t)}_{\text{Mechanical coupling}} \quad \text{(eq. 7)}$$

©EDA Publishing/DTIP 2007                                          ISBN: 978-2-35500-000-3

The gyro needs an oscillator circuit to maintain the drive mode at resonance and a detection circuit to obtain a voltage proportional to the angular velocity Ω. A first stage of preamplifiers converts charges $q_{y+}$ and $q_{y-}$ into voltages. Then a differential stage follows to cancel the capacitive coupling which is in common mode on the two detection electrodes, and a demodulation stage is used to remove the mechanical coupling signal which is in phase quadrature with the Coriolis signal.

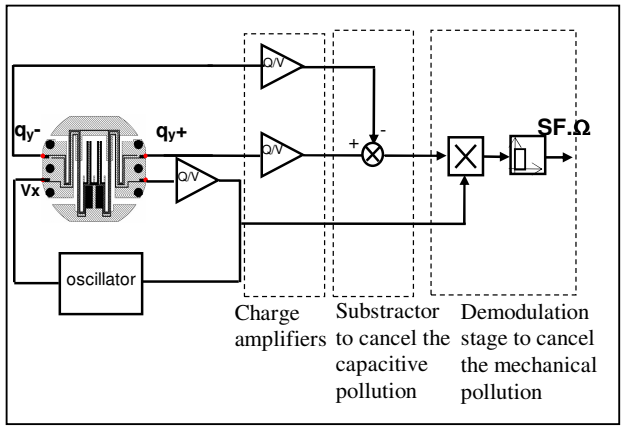

Figure 2: The gyrometer's associated electronics

### 2.2. The preamplifier's specifications

A model of the micro-gyro including its mechanical part, piezoelectricity and the associated electronics has been developed for the gyrometer VIG developed at ONERA [15]. Thanks to simulations performed with this model, it appears that the signal to noise ratio of the preamplifier is proportional to the resolution of the gyro, and the gain stability of the preamplifier limits the stability of the gyrometer's output. It is then important to develop the right preamplifier adapted to the vibrating micro-gyrometer performing low noise and good gain stability.

In the next section, five kinds of preamplifiers are presented and compared in terms of noise and thermal stability.

### 3. COMPARISON OF THE PREAMPLIFIERS

The five preamplifiers are shown on figure 3: the current preamplifier (a) that converts current i into voltage Vs with a feedback resistor, the charge amplifier (b) that converts charge q into voltage Vs with a feedback capacitance, a feedback resistance is used to discharge the capacitance in low frequency and make the circuit stable, the voltage amplifier (c), the differential charge amplifier (d), and the switched capacitor ( SC ) amplifier (e), the switches are used to discharge the feedback capacitance in low frequency.

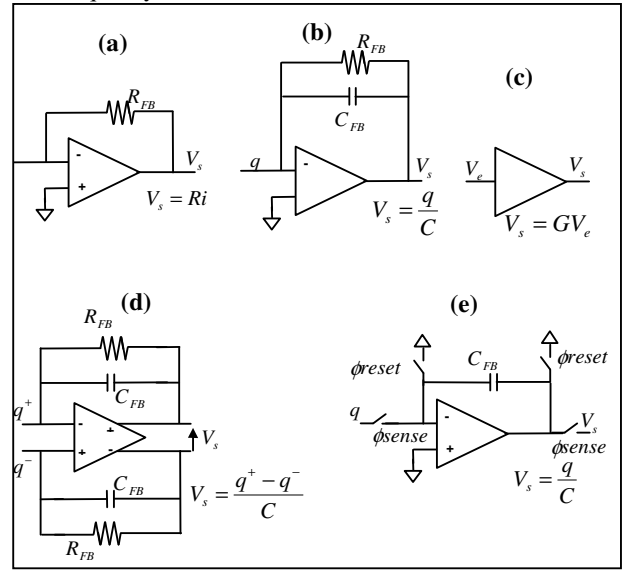

Figure 3: the five kinds of preamplifiers: the current amplifier (a), the charge amplifier (b), the voltage amplifier (c), the differential charge amplifier (d), the switched capacitor charge amplifier (e)

### 3.1. Noise performance

In order to compare the preamplifiers noise performances, we have calculated the charge input referred noise in C/√Hz:

- For the current amplifier :

$$q_B = \sqrt{\frac{4kT}{R_{FB}//R_o}\frac{1}{\omega} + \frac{i_n}{\omega} + e_n C_0 + \frac{e_n}{R_{FB}\omega}}$$

Where $R_{FB}$ is the feedback resistance, Ro Co and ω are respectively the motional resistance, inter-electrodes capacitance, and resonance frequency of the resonator. $e_n$ and $i_n$ are the voltage and current noise of the operational amplifier.

- For the voltage amplifier :

$$q_B = \sqrt{\frac{4kT}{R}\frac{1}{\omega} + \frac{i_n}{\omega} + e_n C_0}$$

*R. Levy[1], A.Dupret[1], H. Mathias[1], J-P. Gilles[1], Serge Muller[2], Bruno Eisenbeis[2], S. Megherbi[1]*
*NOISE AND THERMAL STABILITY OF VIBRATING MICRO-GYROMETERS PREAMPLIFIERS*

R is a resistance put in parallel with Co to discharge Co in low frequency to make the circuit stable.

- For the charge amplifier and the differential charge amplifier:

$$q_B = \sqrt{\frac{4kT}{R_{FB} // Ro}\frac{1}{\omega} + \frac{i_n}{\omega} + e_n(C_0 + C_{FB})}$$

- For the SC amplifier, the largest noise is the kt/C noise due to charge injection of the switches.

Considering the gyrometer VIG, Co=1pF, Ro=1,5 MΩ, and $\omega$=2.10$^5$ rad, the OpAmp voltage noise is the largest noise contribution for the preamplifiers without switches. As $C_{FB} = \frac{1}{R_{FB}\omega}$ to obtain the same output voltage Vs, the noises of the charge and current amplifiers are equal. The noise of the voltage amplifier is smaller.

### 3.2. Thermal stability

The preamplifier gain is set :

- by the amplifier's gain and the inter-electrodes capacitance of the resonator for the voltage amplifier:

$$V_s = G\frac{q^+ - q^-}{C_0}$$

A major drawback of the voltage amplifier is its sensitivity to input parasitic capacitances Cp that makes it unstable over temperature whereas the charge and current amplifiers are not sensitive to input parasitic capacitances because their input voltage is put at ground.

$$V_s = G\frac{q^+ - q^-}{C_0 + C_p}$$

- by the feedback resistance for the current amplifier:
$$V_s = R_{FB}\omega(q^+ - q^-)$$

Frequency drifts over temperature make the current preamplifier drift.

- by the feedback capacitors for the charge amplifiers (charge amplifier, differential charge amplifier and SC charge amplifier):

$$V_s = \frac{q^+ - q^-}{C_{FB}}$$

We conclude that charge amplifiers are the best suited amplifiers to achieve gyrometers thermal stability.

The capacitance and resistance thermal stabilities are, for SMD components 30 ppm/°C. In order to increase the differential gain after the subtractor, the two feedback capacitances can be matched to have the same thermal variations.

### 3.3. Integration

The current amplifier needs a high value resistance for the feedback resistor $R_{CR}$ =10 MΩ. This resistance is too high to be integrated and would take too much space on the chip.

For the charge amplifiers, the feedback resistor doesn't have to be neither stable nor linear because it is the feedback capacitance that sets the gain. It is then possible to use a transistor as a nonlinear resistance. This way the charge amplifiers can be integrated.

The voltage amplifier can also be integrated.

### 3.4 CONCLUSION

The benefits and drawbacks of the five preamplifiers are shown on table 1.

The charge preamplifiers are the most stable preamplifiers over temperature. The switch capacitor charge amplifier doesn't show a good signal to noise ratio because of charge injection. The differential amplifier has the same signal to noise ratio that the charge amplifier and has a better common mode rejection due to its differential architecture.

|  | Signal /noise ratio | Thermal stability | integration | simplicity |
|---|---|---|---|---|
| Charge amplifier | + | + | + | + |
| SC charge amplifier | - | + | + | - |
| Differential charge amplifier | + | ++ | + | - |
| Current amplifier | + | - | - | + |
| Voltage amplifier | + | - | + | + |

Table 1: Benefits and drawbacks of the five preamplifiers

*R. Levy[1], A.Dupret[1], H. Mathias[1], J-P. Gilles[1], Serge Muller[2], Bruno Eisenbeis[2], S. Megherbi[1]*
*NOISE AND THERMAL STABILITY OF VIBRATING MICRO-GYROMETERS PREAMPLIFIERS*

Without taking into account the simplicity of the design, the best suited preamplifier for the micro-gyrometer VIG is the differential charge amplifier. Experimental results for the charge amplifier are shown in the next section. In order to increase the thermal stability by components matching, an integrated circuit design of the differential charge amplifier is presented in section 5.

## 4. EXPERIMENTAL RESULTS WITH THE CHARGE AMPLIFIER

In order to validate the noise and thermal stability calculations, experiments are performed with the charge amplifier connected to a VIG gyro.
The output noise measured (figure 4) of 25 nV/√Hz at 30kHz is in good agreement with the theoretical calculation.
The output amplitude thermal drift (figure 5) of 1mV in the temperature range from -40°C to 80°C is also in good agreement with the theoretical calculation.

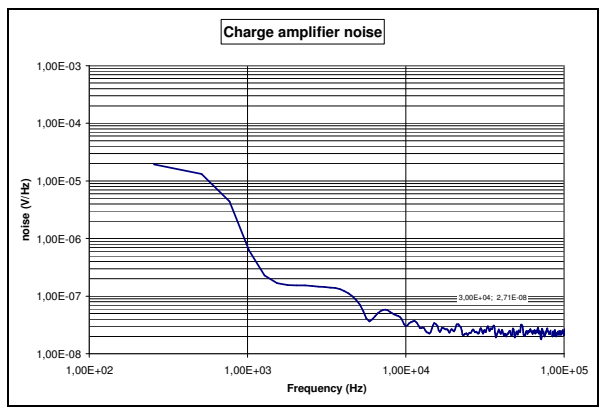

Figure 4: charge amplifier output noise vs frequency

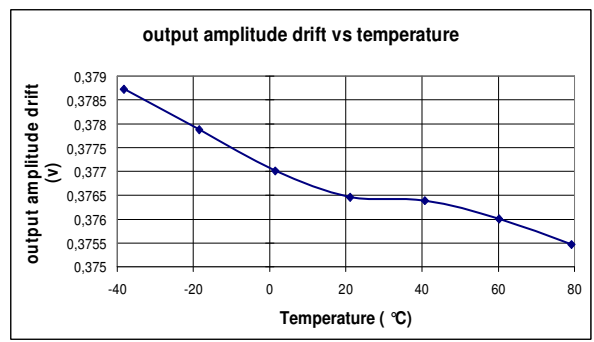

Figure 5: charge amplifier output amplitude drift vs temperature

## 5. INTEGRATED PREAMPLIFIER DESIGN

### 5.1. Low noise design

A specific design is developed to achieve low noise at the resonator frequency (30kHz for the VIG); the input transistors have very big W/L ratios: W/L = 2200. The input referred noise obtained is: $e_n$=5nV/√Hz.

### 5.2. Low thermal drift design

In order to obtain a low thermal drift, the two feedback capacitances that set the differential preamplifier gain are matched. The biasing stage is a bandgap circuit designed to achieve a low thermal sensitivity in the temperature range from -40°C to 80°C.

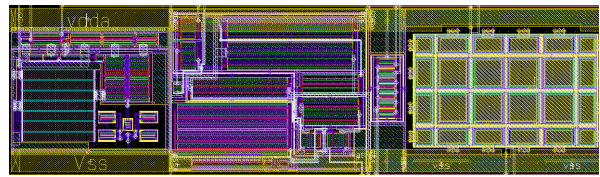

Figure 6: layout of the integrated differential charge amplifier: on the left: the bandgap circuit, on the middle the low noise differential OpAmp, on the right: the two matched feedback capacitors.

## CONCLUSION

Five kinds of preamplifiers for micro-gyrometers have been presented and compared. The switched capacitor amplifier has the largest noise, the current amplifier and voltage amplifiers show gain drifts over temperature. The differential charge amplifier shows a good thermal stability and low noise

In order to increase the differential charge amplifier thermal stability, its integration is finally presented with special care on noise reduction, and feedback capacitor matching. This integrated circuit has been developed and is being realized. The next step is to test the IC with the VIG gyro.

## REFERENCES

[1] O. Le Traon & al, "The VIA vibrating beam accelerometer: a new quartz micromachined sensor", *Proceedings of the Frequency and Time Forum,* vol. 2, *pp 1041-1044*, 1999

[2] M.Lemkin & al , "a three-axis micromachined accelerometer with a CMOS position-sense interface and

digital offset-trim electronics", *IEEE J.of Solid State Circuits*, vol. 34, No 4, *pp. 456-468*, 1999

[3] F. Ayazi & al, "A HARPSS polysilicon vibrating ring gyroscope", Journal of MEMS, vol.10 No 2, 2001.

[4] W. Geiger & al, "The silicon angular rate sebnsor system DAVED", *sensors and actuactors 84, pp 280-284*, 2000.

[5] Il-Han Hwang & al , "Self-actuating biosensor using a piezoelectric cantilever and its optimization", *Journal of Physics: conference series 34, pp 362-367*, 2006.

[6] David-A Mendels & al , "Dynamic propertie of AFM cantilevers and the calibration of their spring constants", *Journal of Micromechanics and Microengineering, pp 1720-1733*, 2006.

[7] R. Levy & al, "A new analog oscillator electronics applied to a piezoelectric vibrating gyro", *Proceedings of the Ultrasonics, Ferroelectrics and Frequency Control conference,* 2004

[8] D. Fang & al, "A Low-Power Low-Noise Capacitive Sensing Amplifier for Integrated CMOS-MEMS Inertial Sensors", *Proceedings of the Circuits, Signals and Systems conference*, 2004

[9] M. Suster & al, **"**Low-noise CMOS integrated sensing electronics for capacitive MEMS strain sensors"*, Proceedings of the IEEE Custom Integrated Circuits Conference**, pp. 693-696*, 2004.

[10] B. Valiki Amini & al, "Micro-gravity capacitive silicon-on-insulator accelerometers", *Journal of Micromechanics and Microengineering No 15, pp 2113-2120*, 2005.

[11] N. Yazdi & al, "A Low-Power Interface Circuit for Capacitive Sensors," *Proc. Solid-State Sensors & Actuators Workshop*, pp. 215-218, 1996.

[12] F.Ayazi & al, A harpss polysilicon vibrating ring gyroscope, *Journal of microelectromechanical systems,* vol. 10, n°2, 2001.

[13] D. Janiaud & al , "The VIG Vibrating Integrated Gyrometer: a new quartz micromachined sensor", *Symposium Gyro Technology*, 2003.

[14] W.Geiger & al, The silicon angular rate sensor system DAVED, *sensors and actuators 84, pp 280-284*, 2000.

[15] R.Levy & al, "Study and realization of an electronics adapted to the piezoelectric vibrating micro-gyrometer", *PhD from university Paris 11*, 2001.